\documentclass[twocolumn,aps,showpacs,epsf]{revtex4}
\usepackage{bm}  
\usepackage{graphicx}
\usepackage{amssymb}
\usepackage{color}
\usepackage{amscd}
\usepackage{epsfig}
\def\beq{\begin{equation}}
\def\eeq{\end{equation}}
\def\beqa{\begin{eqnarray}}
\def\eeqa{\end{eqnarray}}

\begin{document}

\title{Hydrodynamics of self-propelled hard rods}
\author{Aparna Baskaran and M. Cristina Marchetti}
\address{Physics Department, Syracuse University, Syracuse NY 13244}
\date{\today}

\begin{abstract}
Motivated by recent simulations and by experiments on aggregation
of gliding bacteria, we study a model of the collective dynamics
of self-propelled hard rods on a substrate in two dimensions. The
rods have finite size, interact via excluded volume and their
dynamics is overdamped by the interaction with the substrate.
Starting from a microscopic model with non-thermal noise sources,
a continuum description of the system is derived. The hydrodynamic
equations are then used to characterize the possible steady states
of the systems and their stability as a function of the particles
packing fraction and the speed of self propulsion.
\end{abstract}

\maketitle

\section{Introduction}
Self propelled particles draw energy from internal or external
sources and dissipate this energy by moving through the medium
they inhabit. Such a definition encompasses a wide class of
systems such as fish schools, bacterial colonies and monolayers of
vibrated granular rods. In all of these the energy input that
maintains the system out of equilibrium is on each unit, rather
than at the boundaries as in more conventional non-equilibrium
situations~\cite{cross_review}. A striking phenomenon exhibited by
these systems is flocking, the emergence of a coherently moving
body of self propelled entities of size much larger than the
length scale of the inter-particle interaction.

Extensive theoretical effort has been devoted to understanding
these non-equilibrium systems. Two distinct approaches have been
used. First, starting with the seminal work by Vicsek
\cite{Vicsek}, a number of numerical studies of simplified
rule-based model systems have been carried out. A second approach
has been to write down generic continuum theories based on
symmetry considerations \cite{TonerRev}. These methods are well
developed for equilibrium systems, where general properties such
as the fluctuation-dissipation theorem yield strong constraints on
the parameters of the dynamical model
\cite{ChaikinLubensky,deGrootMazur}.  They are also very useful
for non-equilibrium systems, although in this case the parameters
in the equations remain largely undetermined. Both approaches have
exposed the dramatically different nature of order-disorder
transitions and of the fluctuations in the various phases
exhibited by self-propelled systems when compared to their
equilibrium counterparts. Collections of self-propelled units can
exhibit long range order in two dimensions \cite{TonerRev}, in
sharp contrast to equilibrium systems, where the continuous
rotational symmetry cannot spontaneously be broken in systems with
short-range interactions~\cite{Lubensky_book}.  Recent numerical
work has shown that the Viscek model and other closely related
models with polar aligning interactions exhibit a first order
(discontinuous) transition from a disordered to an ordered state
in two dimensions when the density and the noise amplitude in the
system are varied \cite{Chate1}. On the other hand, when this
model is modified to induce strictly nematic order in the system,
the phase transition becomes continuous \cite {Chate2}. In
addition, large number fluctuations have been predicted and
observed in the homogeneous states of these driven systems
\cite{RamaswamyPRL,RamaswamyEPL,RamaswamyJSTAT}.

Non-equilibrium statistical mechanics is a powerful tool that can
be brought to bear to describe this rich class of systems. One class of questions pertains to the
identification of the underlying collective mechanisms that give
rise to emergent behavior. Comparison of
theoretical and  numerical studies of model systems can provide
insight into the physical origin of the various phenomena captured
by the generic continuum theories. The second class of questions
is associated with identifying minimal microscopic models capable
of accounting for a given observed phenomenon. This is important
when trying to generalize concepts developed in the context of
animal group behavior to artificial systems considered in
collective robotics \cite{AdBh}.
An example of this approach is the recent work by Bertin et al.
\cite{BoltzPap} who used non-equilibrium statistical mechanics to
derive continuum hydrodynamic equations for the polar Viscek
model, with parameters given explicitly in terms of parameters of
the microscopic model.

In this work, we study the collective behavior of self propelled
hard rods with excluded volume interactions, moving on a substrate
in two dimensions.  Part of our motivation comes from recent
simulations of this systems by Peruani et al. \cite{Peruani06}.
Unlike earlier work~\cite{BoltzPap,Aranson}, our theory accounts
for the extended shape of the self-propelled units. We consider a
minimal model, with the aim of identifying the simplest physical
mechanisms that can lead to self-organization in self-propelled
systems with short-range interactions.  The non-equilibrium
statistical mechanics for this model  is developed systematically
and results in coarse-grained hydrodynamic equations for the slow
variables of the system. The macroscopic equations are then used
to characterize the homogeneous steady states and their stability
as a function of particle density and self propulsion velocity.
This analysis examines the effect of flow from self propulsion on
the intrinsic entropic order that arises from the finite size of
the rods. Our work provides a  microscopic basis for some of the
phenomena predicted in active nematic via generic continuum
theories \cite{RamaswamyEPL} and observed in simulations
\cite{Peruani06,Chate2,MR06} .

A real system that may be described by our minimal model  is
myxobacteria on a substrate under starvation conditions. These
bacteria are rod-like in shape, with an aspect ratio of  about 7,
and retain their shape under movement. Under starvation conditions
the C-signalling mechanism in these bacteria that give rise to
reversal of the direction of motion is suppressed and alignment is
thought to be brought about by steric interactions
\cite{Igoshin1}.

The layout of the paper is as follows. First, the microscopic
model is introduced and the non equilibrium statistical mechanics
developed. In a low density approximation, this is described by a
Smoluchowski equation. From this starting point, the hydrodynamic
equations associated with the slow variables of the system, namely
the density, polarization (related to the self propulsion flow
field) and nematic order parameter, are derived. Next, the
homogeneous steady states  are identified and their linear
stability with respect to spatial fluctuations characterized. This
analysis is used to interpret the physical mechanism behind the
emergence of inhomogeneities in the nematic state observed in the
numerical work of Ref. \cite{Peruani06} and the large number
fluctuations predicted in \cite{RamaswamyEPL}.

\section{Microscopic model}

As a minimal physical model for a system of self propelled
particles, we consider a collection of rigid rods of length
$\ell>>b$, with $b$ the rod diameter, moving on a substrate in two
dimensions ($2d$).  Each rod is ``self-propelled" in that it has a
force $F$ acting on its center of mass and oriented along its long
axis. The rods interact only via excluded volume interaction and
their dynamics is overdamped. The exchange of momentum with the
substrate due to friction provides the physical mechanism that
results in all the rods on average having a constant velocity of
magnitude $v_0$ along their long axis. At high packing fractions,
the excluded volume interaction gives rise to orientational order,
as in passive nematic liquid crystals.

Note that a flock of self propelled particles is characterized by
three properties : 1) centering - the ability of a collection of
such particles to stay together, 2) velocity matching - all the
particles are on average characterized by the same velocity, and
3) alignment - the particles are asymmetric and exhibit ordered
states with either nematic or polar symmetry \cite{Reynolds}. In
our model, the rods have no attractive interactions and hence will
not form a cohesive flock. We assume that the system is at a fixed
homogeneous density $\rho_0$, which is maintained by confinement
at the boundary of the system. Alignment arises from the excluded
volume interaction. As noted above, the force $F$ together with
friction from the substrate gives rise to velocity matching among
the particles. Hence the minimal model considered here has the
necessary ingredients to study the collective behavior of
self-propelled particles.

Each rod is characterized by the position $\mathbf{r}_{i}$ of its
center of mass  and its orientation with respect to an external
axis, described by a unit vector
$\widehat{\mathbf{u}}_{i}=\big(\cos \theta _{i},\sin \theta
_{i}\big)$ directed along the axis of the rod. The microscopic
dynamics of the $i$-th rod is controlled by  coupled Langevin
equations,
\begin{equation}
\frac{\partial r_{i\alpha }}{\partial t}=-D^i_{\alpha \beta
}\sum_{j}\frac{\partial }{\partial r_{i\beta }} V_{ex}\left( {\bf
r}_{ij},\theta _{i},\theta _{j}\right) +v_{0}\widehat{u} _{i\alpha
}+\eta _{{i\alpha }}\left( t\right) \;,  \label{1.1}
\end{equation}
\begin{equation}
\frac{\partial \theta _{i}}{\partial t}=-D_{R}\sum_{j}
\frac{\partial }{\partial \theta _{i}}V_{ex}\left( {\bf
r}_{ij},\theta _{i},\theta _{j}\right) +\eta^R _{{i}}\left(
t\right) \;, \label{1.2}
\end{equation}
where Greek indices denote Cartesian components and ${\bf
r}_{ij}={\bf r}_i-{\bf r}_j$. The self-propulsion velocity of
magnitude $v_0$ is directed along
 the long axis of the rod.
We choose coordinates so that the $z$ axis is normal to the plane
of the substrate. The excluded volume interaction $V\left(
r_{ij},\theta _{i},\theta _{j}\right) $ has the Onsager form, with
\begin{eqnarray}
V_{ex}\left( {\bf r}_{ij},\theta _{i},\theta _{j}\right)
&=&1\text{ \ \ \ \ \ \ \
if particles }i,j\text{ intersect}  \nonumber \\
&=&0\text{ \ \ \ \ \ \ \ otherwise.}  \label{1.3}
\end{eqnarray}
This effective interaction can be derived from the collision rules
that govern the dynamics of $2d$ hard rods and represents a very
good approximation to the true momentum exchange in the overdamped
limit~\cite {aparna2}. The diffusion tensor, $D^i_{\alpha \beta
}$, and the rotational diffusion rate, $D_{R}$, are determined by
the energy scale associated with the excluded volume interaction
and the translational and rotational damping provided by the
substrate. Finally, $\eta _{{i\alpha }}$ and $\eta^R _{{i}} $ are
white noise sources with correlations
\begin{eqnarray}
\left\langle \eta_{{i\alpha }}\left( t\right) \eta _{{j\beta
}}\left( t^{\prime }\right) \right\rangle &=&\tilde{D}^i_{\alpha
\beta }\delta _{ij}\delta \left( t-t^{\prime }\right)\;,
\label{1.5}\\
\left\langle \eta ^R_{{i}}\left( t\right) \eta ^R_{{j}}\left(
t\right) \right\rangle &=&\tilde{D}_{R}\delta _{ij}\delta \left(
t-t^{\prime }\right) \;.  \label{1.6}
\end{eqnarray}
For systems in thermal equilibrium the fluctuation-dissipation
theorem requires that the amplitudes of the correlation of the
thermal noise be identical to the diffusive parameters controlling
the excluded volume interaction, i.e.,
$\tilde{D}^i_{\alpha\beta}=D^i_{\alpha\beta}$ and
$\tilde{D}_{R}=D_R$. This identification does not hold  for
non-equilibrium systems such as the one considered here, where the
noise has non-thermal contributions. For self-propelled
suspensions noise arises in general from thermal Brownian motion,
hydrodynamic interactions among the self-propelled units, and
intrinsic fluctuations in the activity of each self-propelled
unit. Brownian noise is expected to be very small for particles
such as bacteria of size larger than a few microns. The other two
sources of noise are intrinsically non-thermal and will generally
not obey a fluctuation-dissipation theorem. It can be shown,
however,  that the equality still holds for a dilute solution
 in the regime of small self propulsion velocities \cite
{aparna2}. When the self propulsion becomes large the relationship
between these diffusion constants is complicated. In the
subsequent analysis, the emergent physics of the system is
independent of these details. Therefore for simplicity we assume
the noise amplitudes are simply given by their equilibrium
counterparts, with
\begin{equation}
\tilde{D}^i_{\alpha \beta }=D^i_{\alpha \beta }
=D_{\parallel}\widehat{u}_{i\alpha }\widehat{u}_{i\beta }
+D_\perp\left( \delta_{\alpha\beta}-\widehat{u}_{i\alpha
}\widehat{u}_{i\beta } \right) \;,.\label{1.7}
\end{equation}
For long thin rods the longitudinal and transverse diffusion
constants $D_\parallel$ and $D_\perp$ are simply related as
$D_\parallel=2D_\perp\equiv 2 D_0$, with $D_0=k_{B}T/\zeta$ for a
single rod at low density where $T$ is the temperature  and
$\zeta$ is a material dependent friction constant characterizing
the interaction of the rod and substrate. A similar approximation
is used for the rotational noise amplitude, with
\begin{equation}
\tilde{D}_{R}=D_{R}\simeq\frac{6D_0}{\ell^{2}}\;, \label{1.8}
\end{equation}
where the last equality holds for infinitely thin rods at low
density.

\begin{figure}
\centerline{\epsfxsize=6cm \epsfbox{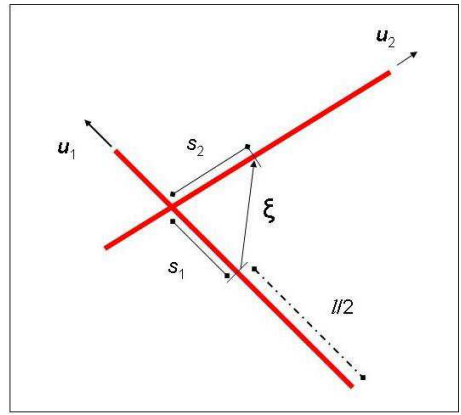}}
\caption{Geometry of overlap between two rods of length $\ell$.
Here $\bm\xi ={\bf r}_2-{\bf r}_1$ is the separation of the
centers of mass of the two rods.}\label{rods}
\end{figure}

The noise averaged statistical mechanics of a system of overdamped
Langevin equations is given by a Smoluchowski equation for the
$N$-particle phase space density. In the low density limit, when
two particle correlations can be neglected, this results in an
effective mean field Smoluchowski equation of the familiar form
for the one-particle probability density,
$c\left(\mathbf{r},\theta ,t\right) $, representing the density of
rods with center of mass at $\mathbf{r}$ and orientation
 $\theta $ at time $t$. Its dynamics is governed by the equation
\begin{equation}
\frac{\partial c\left( \mathbf{r},\theta ,t\right) }{\partial
t}=-\bm\nabla\cdot \mathbf{J}^{T}-\frac{\partial }{\partial \theta
}J^{R}\;, \label{1.S}
\end{equation}
with translational and rotational currents
\begin{eqnarray}
{J}_{\alpha }^{T}&=&-D_{\alpha \beta }~c\frac{\partial V\left(
{\bf r},\theta \right)}{\partial {r}_{\beta }} +v_{0}\widehat{{u}}
_{\beta }c-D_{\alpha \beta }\frac{\partial c}{\partial {r}_{\beta
}}\;, \label{1.S.1}\\
J^{R}&=&-D_{R}c\frac{\partial V\left( {\bf r},\theta \right)
}{\partial \theta }-D_{R}\frac{\partial c}{\partial \theta }\;.
\label{1.S.2}
\end{eqnarray}
 In Eqs. (\ref{1.S.1}) and (\ref{1.S.2})  the two-particle Onsager interaction
defined in Eq.~(\ref{1.3}) enters in its mean field form
\begin{eqnarray}
V\left( {\bf r}_1,\theta_1 \right) &=&\int d\mathbf{r}_{2}\int
d\theta _{2}V_{ex}\left( \mathbf{r}_{12},\theta _{1},\theta
_{2}\right) c\left( \mathbf{r}_{2},\theta
_{2},t\right)  \nonumber \\
&=&\int d\bm{\xi }\int d\theta _{2}\left| \widehat{\bf
u}_{1}\times \widehat{\bf u} _{2}\right| c\left(
\mathbf{r}_{1}+\bm{\xi },\theta _{2},t\right)\; \label{1.9}
\end{eqnarray}
where $\bm\xi ={\bf r}_2-{\bf r}_1\simeq
s_{1}\mathbf{\widehat{u}}_{1}-s_{2}\mathbf{\widehat{u}}_{2}$ and
$-\ell/2\leq s_i\leq \ell/2$ parametrizes the position along the
$i$-th rod measured from its center of mass. As shown in
Fig.~\ref{rods}, this is  a measure of the area excluded by a rod
at a point $\mathbf{r}_1$, oriented in the direction $\theta_1 $
to all other rods in the system.

We emphasize that the self propulsion mechanism enters the
Langevin equations (\ref{1.1}) and (\ref{1.2}) only in the form of
a center of mass force. Therefore it gives rise to no microscopic
torques. This is reflected in the Smoluchowski equation above in
the fact that the rotational fluxes in Eq.~(\ref{1.S.2}) contain
no information about the self propulsion. The velocity $v_{0}$
enters only the translational flux as an additional mass flux
along the orientation of the rod. In the absence of self
propulsion ($v_0=0$), the microscopic equations and hence the
Smoluchowski equation are both invariant when
$\widehat{\mathbf{u}} _{i}\rightarrow -\widehat{\mathbf{u}}_{i}$
for all rods and the system cannot exhibit a macroscopic polar
state. When $v_0\not=0$, this symmetry is broken at the level of
the Langevin equations. One might then expect that the system can
order in a uniform polar state. But since bulk states are
controlled entirely by the rotational fluxes which still possess
the nematic symmetry, such a homogeneous polar state cannot exist.
This is not the consequence of any of the approximations
 made to obtain the Smoluchowski equation, but is a true
property of the non-equilibrium model system described by the
microscopic dynamics given in Eqs.~(\ref{1.1}) and (\ref{1.2}).
The self propulsion term does, however, affect the fluctuations
the systems, as described in the next
 section.

\section{Hydrodynamic Equations}

In this section we obtain coarse grained equations appropriate for
describing the dynamics of the system on length scales large
compared to the length of the rods and  time scales long compared
to the microscopic diffusion times in the Langevin equations. In
this limit we expect that the dynamics will be controlled by the
conserved densities  and the fields associated with possible
broken symmetries. For an overdamped system of the type considered
here the only conserved quantity is the concentration of rods. It
can be defined as the zeroth moment of the probability
distribution $c\left(\mathbf{r},\widehat{\mathbf{u}} ,t\right) $,
\begin{equation}
\rho \left( \mathbf{r},t\right) =\int d\widehat{\mathbf{u}}~c\left(
\mathbf{r} ,\widehat{\mathbf{u}},t\right).  \label{2.1}
\end{equation}
Both polar and nematic order are in principle possible in a gas of
self-propelled rods. These can be described by a polarization vector $%
\mathbf{P}(\mathbf{r},t)$ and the nematic alignment tensor $Q_{\alpha\beta}(%
\mathbf{r},t)$ defined as first and second moments, respectively,
of $c\left(\mathbf{r},\widehat{\mathbf{u}} ,t\right) $,
\begin{equation}
\rho \left( \mathbf{r},t\right) \mathbf{P}\left( \mathbf{r},t\right) =\int d%
\widehat{\mathbf{u}}~ \widehat{\mathbf{u}}~c\left(
\mathbf{r},\widehat{\mathbf{u}},t\right), \label{2.2}
\end{equation}
\begin{equation}
\rho \left( \mathbf{r},t\right) Q_{\alpha \beta }\left(
\mathbf{r},t\right) =\int d\widehat{\mathbf{u}} \left(
\widehat{u}_{\alpha }\widehat{u}_{\beta }- \frac{1}{2}\delta
_{\alpha \beta }\right) c\left( \mathbf{r},\widehat{
\mathbf{u}},t\right) .  \label{2.3}
\end{equation}
Since each rod has a self propulsion velocity
$v_{0}\widehat{\mathbf{u}}$, the polarization is also proportional
to the self propulsion flow field.

The equations for these continuum fields are obtained by taking
the corresponding moments of the Smoluchowski equation (\ref{1.S})
and have the form
\begin{equation}
\partial _{t}\rho =-\bm\nabla\cdot\mathbf{J}\;,  \label{2.4}
\end{equation}
\begin{equation}
\partial _{t}\big(\rho P_{\alpha }\big)=-\frac{\partial }{\partial r_{\beta }}
J_{\alpha \beta }-R_{\alpha }  \label{2.5}
\end{equation}
\begin{equation}
\partial _{t}\big(\rho Q_{\alpha \beta }\big)=-\frac{\partial }{\partial r_{\gamma }}
J_{\alpha \beta \gamma }-R_{\alpha \beta }  \label{2.6}
\end{equation}
where
\begin{equation}
\left(
\begin{array}{c}
J_{\gamma } \\
J_{\alpha \gamma } \\
J_{\alpha \beta \gamma }
\end{array}
\right) =\int d\widehat{\mathbf{u}}\left(
\begin{array}{c}
1 \\
\widehat{u}_{\alpha } \\
\left( \widehat{u}_{\alpha }\widehat{u}_{\beta }-\frac{1}{2}\delta
_{\alpha \beta }\right)
\end{array}
\right) J_{\gamma }^{T}\;,  \label{2.7}
\end{equation}
and
\begin{equation}
\left(
\begin{array}{c}
R_{\alpha } \\
R_{\alpha \beta }
\end{array}
\right) =\int d\widehat{\mathbf{u}}\left(
\begin{array}{c}
\widehat{u}_{\alpha } \\
\left( \widehat{u}_{\alpha }\widehat{u}_{\beta }-\frac{1}{2}\delta
_{\alpha \beta }\right)
\end{array}
\right) \frac{\partial J^{R}}{\partial\theta}\;, \label{2.8}
\end{equation}
with the fluxes $\mathbf{J}^{T}$ and $J^{R}$ as given in
Eqs.~(\ref{1.S.1}) and (\ref{1.S.2}). In order to obtain a closed
set of hydrodynamic equations, we need to evaluate the
constitutive equations for the fluxes on the right hand side of
Eqs.~(\ref{2.4}-\ref{2.6}). This requires various approximations.
First, since we are interested in a long-wavelength description of
the system, the nonlocal dependence on the concentration field in
Eq.~(\ref{1.9}) is expanded in gradients as
\begin{eqnarray}
c(\mathbf{r}_{1}+\bm{\xi },&&\mathbf{\widehat{u}}_{2},t) =c\left(
\mathbf{r}_{1},\mathbf{\widehat{u}}_{2}\right) +\xi _{\alpha
}\partial _{r_{1\alpha
}}c\left( \mathbf{r}_{1},\mathbf{\widehat{u}}_{2}\right)  \nonumber \\
&&+\frac{1}{2 }\xi _{\alpha }\xi _{\beta }\partial _{r_{1\alpha
}}\partial _{r_{1\beta }}c\left(
\mathbf{r}_{1},\mathbf{\widehat{u}}_{2}\right) +O\left( \nabla
^{3}\right) \;,  \label{2.9}
\end{eqnarray}
where the expansion has been truncated at second order in the
gradients. This amounts to a hydrodynamic description up through
Navier-Stokes order in the coarse grained densities.

Secondly, the balance equations for the polarization and the
nematic order parameter couple to higher moments of the
probability distribution. To obtain a closed set of equations we
need an additional assumption that allows these higher moments to
be expressed in terms of the first three moments of the
probability distribution. We assume that at long times (times
longer than all microscopic diffusion times) the higher moments
become functionals of the first three moments \cite{McLBook} and
the probability distribution has the form
\begin{equation}
c\left( \mathbf{r},\widehat{\mathbf{u}},t\right) \rightarrow
c\left( \widehat{\mathbf{u}},\left[ y_{\alpha }\left(
\mathbf{r},t\right) \right] \right)\;,  \label{2.10}
\end{equation}
where $y_{\alpha }\left( \mathbf{r},t\right) = \left\{ \rho \left(
\mathbf{r},t\right) ,\mathbf{P}\left( \mathbf{r} ,t\right)
,Q_{\alpha \beta }\left( \mathbf{r},t\right) \right\} $ and the
square brackets denote the fact that the probability distribution
$c$ is a functional of these fields. Once such a functional
assumption is made, the Smoluchowski equation can be self
consistently solved to obtain the probability distribution in
terms of the slow variables in the system and their gradients. In
the context of equilibrium systems, this procedure is implemented
using slow variables that are hydrodynamic, i.e., they are the
conserved quantities of the system. But as has been shown in the
context of granular fluids, it can carried out using overdamped
variables as in the case at hand as well \cite{BDKS}.

Here this procedure is implemented in a much simpler context. It
is assumed that the probability distribution is a linear
functional of the slow variables and the solution to the
Smoluchowski equation is obtained to lowest order in the gradients
in the slow variables of interest here. Then, it follows
immediately that the probability distribution is given by
\begin{eqnarray}
c\left( \mathbf{r},\widehat{\mathbf{u}},t\right) &=&\frac{1}{2\pi
}\rho
\left( \mathbf{r},t\right) \Big[ 1+2\mathbf{P}\left( \mathbf{r}%
,t\right)\cdot \widehat{\mathbf{u}}  \nonumber \\
&&+4Q_{\alpha \beta }\left( \widehat{u}_{\alpha }\widehat{u}
_{\beta }-\frac{ 1}{2}\delta _{\alpha \beta }\right) \Big] \;.
\label{2.11}
\end{eqnarray}
Using the two approximations given by Eq.~(\ref{2.9}) and
Eq.~(\ref{2.11}), a set of closed macroscopic equations for the
density, polarization and nematic order parameter are obtained.
Including only diffusion and self propulsion contributions to the
fluxes,  these equations are given by
\begin{widetext}
\begin{equation}
\partial _{t}\rho+v_0\bm\nabla\cdot(\rho{\bf P})=\frac{3D_0}{4}\nabla^2\rho+\frac{D_0}{2}\partial _{\alpha }\partial_\beta(\rho Q_{\alpha\beta} ) \label{H.2}
\end{equation}
\begin{equation}
\partial _{t}\big(\rho P_{\alpha }\big)+\frac12 v_0\partial_\alpha\rho
+v_0\partial_\beta(\rho Q_{\alpha\beta})=-D_{R}\rho P_\alpha
+\frac{5D_0}{8}\nabla^2(\rho
P_\alpha)+\frac{D_0}{4}\partial_\alpha\bm\nabla\cdot(\rho{\bf P})
\label{H.3}
\end{equation}
\begin{eqnarray}
\partial _{t}\big(\rho Q_{\alpha \beta }\big)+&&\frac{v_0}{4}\Big[\partial_\alpha(\rho P_\beta)
+\partial_\beta(\rho
P_\alpha)-\delta_{\alpha\beta}\bm\nabla\cdot(\rho{\bf P})\Big]
=-4D_{R}\rho Q_{\alpha \beta } \nonumber\\
&&+\frac{D_0}{8}\Big[\partial_\alpha\partial_\beta-\frac12\delta_{\alpha\beta}
\bm\nabla\cdot\Big]\rho
+\frac{D_0}{6}\Big[\partial_\gamma\big(\partial_\alpha\rho
Q_{\beta\gamma} +\partial_\beta\rho
Q_{\alpha\gamma}-\delta_{\alpha\beta}\partial_\sigma\rho
Q_{\sigma\gamma}\big) +\frac{7}{2}\nabla^2\rho
Q_{\alpha\beta}\Big] \label{H.4}
\end{eqnarray}
\end{widetext}
The  excluded volume contributions to the fluxes are given in
Appendix A. In Eqs.~(\ref{H.2}-\ref{H.4}), self propulsion
generates convective-type terms that couple the density and the
nematic order parameter to the polarization, reflecting the fact
that self propulsion yields a mass flux in the direction of
polarization. In the following sections, these equations are used
to study the possible steady state of the systems and their
stability.

\subsection{Homogenoues  states}

The bulk states of the system are determined by the solutions of
the homogeneous hydrodynamic equations. Dropping all gradients
terms, but including excluded volume effects, these are given by
\begin{equation}
\partial _{t}\rho =0,  \label{2.1.1}
\end{equation}
\begin{equation}
\partial _{t}\rho P_{\alpha }=-D_{R}\rho P_{\alpha }+\frac{4}{3}D_{R}\eta _{0}\rho ^{2}P_{\beta
}Q_{\alpha \beta },  \label{2.1.2}
\end{equation}
\begin{equation}
\partial _{t}\rho Q_{\alpha \beta }=-4D_{R}\rho \Big(1-\eta_0\rho/3\Big)Q_{\alpha \beta }\;,  \label{2.1.3}
\end{equation}
where $\eta _{0}=2\ell^{2}/\pi$ is the excluded volume of a rod.
As anticipated,  the self propulsion velocity $v_0$ does not enter
the homogeneous equations. This is an artifact of our modeling of
self propulsion solely as a center of mass force and hence an
effective center of mass velocity in the overdamped limit in the
microscopic model considered here. The homogeneous equations are
therefore identical to those of an equilibrium collection of
overdamped hard rods. At low density the system forms  an
isotropic liquid state, with $\rho=\rho_0$, ${\bf P}=0$ and
$Q_{\alpha\beta}=0$. The isotropic state becomes unstable for
$\rho_0>\rho _N=3\eta_0=3\pi/2\ell^{2}$ where the coefficient of
$Q_{\alpha\beta}$ on the right hand side of Eq.~(\ref{2.1.3})
changes sign. For $\rho_0>\rho_N$ a collection of hard rods forms
a nematic liquid crystal, characterized by  broken orientational
symmetry along the direction of a unit vector $\widehat{\bf n}$
known as the director. The alignment tensor takes the form
\begin{equation}
Q_{\alpha \beta }=S\left( \widehat{n}_{\alpha }\widehat{n}_{\beta
}-\frac{1}{2} \delta _{\alpha \beta }\right)\;.
\end{equation}
This is the well-known result due to Onsager. In mean-field theory
the isotropic-nematic transition is continuous in two dimensions
\cite{EdDoiBook}. To obtain the value $S$ of the order parameter
one must retain terms cubic in $Q_{\alpha\beta}$ in
Eq.~(\ref{2.1.3}) which arise from higher order correlations
neglected here. These yield a term  $\sim w\rho _{0}^{3}Q_{\gamma
\delta }Q_{\gamma \delta }Q_{\alpha \beta }$, where in the present
context $w$ is a phenomenological parameter, independent of
density.
 One then finds $S=\frac{1}{\rho _{0}}\sqrt{\frac{8}{w}\left( \rho _{0}/\rho
_{N}-1\right) }$ in the nematic state.
Finally, there is no homogeneous polarized state.

\subsection{Fluctuations in the isotropic State}

In this section we examine the dynamics of fluctuations in the
isotropic state for $\rho_0<\rho_N$ to study its stability.
 The only hydrodynamic variable in the isotropic state is the concentration of filaments, $\rho$. Fluctuations in both the polarization and the alignment tensor decay on microscopic time scales. However, while fluctuations in the alignment tensor couple to the density to higher order in the gradients and can be safely neglected, polarization fluctuations can qualitatively change the nature of the hydrodynamic modes at intermediate length scales. We therefore consider the coupled dynamics of fluctuations of the density $\delta\rho({\bf r},t)=\rho({\bf r},t)-\rho_0$ and polarization $\delta {\bf P}({\bf r},t)={\bf P}({\bf r},t)$ from their homogeneous values
 $\rho=\rho_0$ and ${\bf P}=0$.
It is convenient to expand the fluctuations in  Fourier components
\begin{equation}
\delta \widetilde{y}\left( \mathbf{k},t\right) =\int
d\mathbf{r}e^{i\mathbf{k }\cdot \mathbf{r}}\delta y\left(
\mathbf{r},t\right) .  \label{2.2.2}
\end{equation}
and to write the polarization in terms of its components
longitudinal and transverse to $\widehat{\mathbf{k}}$, as
${\bf\tilde{P}}({\bf k},t)=\widehat{\mathbf{k}}\delta P_\parallel
({\bf k},t)+\widehat{\mathbf{k}}_{\bot }\delta P_\perp({\bf
k},t)$, where $\widehat{\mathbf{k}}={\bf k}/|{\bf k}|$ and
$\widehat{\mathbf{k}}_{\bot }=\hat{z}\times \widehat{\mathbf{k}}$
are unit vectors longitudinal and perpendicular to ${\bf k}$. The
transverse component of the polarization  decouples from the
density and will be neglected below. The linearized coupled
equations for fluctuations in the density and longitudinal
polarization, which corresponds to splay deformations of the
polarization field, are given by
\begin{equation}
\partial _{t}\delta \widetilde{\rho }=-Dk^{2}\delta \widetilde{
\rho }+ikv_{0}\rho _{0}\delta \widetilde{P}_{\Vert }
\label{2.2.4}
\end{equation}
\begin{equation}
\partial _{t}\delta \widetilde{P}_{\Vert }=-\big[D_{R}+D_pk^2\big]\delta \widetilde{P}_{\Vert
}+ikv_0 \frac{\delta \widetilde{\rho }}{2\rho _{0}}
 \label{2.2.5}
\end{equation}
where in the low density approximation considered here
\begin{eqnarray}
&&D =\frac{3D_{0}}{4}\left[ 1+\frac{\rho _{0}}{\rho _{IN}}
\right]\;,\\
&&D_p=\frac{7}{8}D_0\;.
\end{eqnarray}
Note that the splay diffusion constant $D_p$ can naturally be
written as $D_p=K_1/\zeta$, with $K_1$  the splay elastic constant
and  $\zeta =\frac{k_{B}T}{D_{0}}$ a friction coefficient. At low
density we obtain $K_1=\frac{7}{8}k_BT$.

If $v_0=0$, Eqs.~(\ref{2.2.4}) and (\ref{2.2.5}) decouple: density
fluctuations decay via diffusion and polarization fluctuations are
overdamped with decay rate $D_R$. Conversely, at finite $v_0$, for
$t>>D_R^{-1}$ and long wavelengths, we can assume that
polarization fluctuations relax quickly to a value determined by
the inhomogeneous density field, $\delta \tilde P_\parallel\simeq
ikv_0\delta\tilde\rho/(2D_R\rho_0)$, which yields  traveling
density waves of speed $v_0$.

\begin{figure}
\centerline{\epsfxsize=6cm \epsfbox{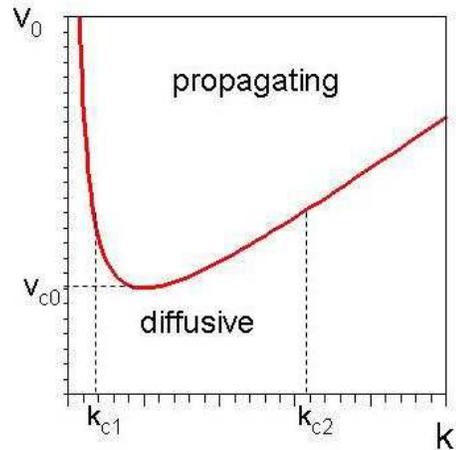}} \caption{A
qualitative description of the crossover from diffusive to
propagating density fluctuations. The solid line (red online) is
the boundary for $v_0(k)$ given in Eq.~(\ref{vck}). For a fixed
value $v_0>v_{c0}$ the modes are propagating in a range of
wavevector $k_{c1}\leq k\leq k_{c2}$.}\label{sound}
\end{figure}

In general we look for wavelike solutions with time dependence
given as linear combinations of terms of the form $\sim
e^{\lambda_\nu(k)t}$, where $\lambda_\nu(k)$ are the dispersion
relations of the hydrodynamic modes of the system. The modes
controlling the decay of density and splay fluctuations in the
isotropic state are given by
\begin{eqnarray}\label{eigenv}
\lambda _{1,2}&=&-\frac{D_{R}+\left(D_p+D\right)
k^{2}}{2} \nonumber\\
&&\pm \frac12\sqrt{\left[ D_{R}+\left( D_p-D\right) k^{2}\right]
^{2}-2v_{0}^{2}k^{2}}\;,
\end{eqnarray}
%
%
%
%
%
%
The sign of the real part of the eigenvalues $\lambda_\nu(k)$
controls the linear stability of the homogeneous state. It is easy
to see that in the isotropic state $Re[\lambda_\nu(k)]<0$ for
$\nu=1,2$ and all values of parameters, indicating that the
isotropic state is linearly stable for all $\rho_0<\rho_N$. The
nature of the modes changes, however, from diffusive  to
propagating when the argument of the square root on the right hand
side of Eq.~(\ref{eigenv}) changes sign. At low density $D_p>D$
and the system exhibits propagating density waves for $v_0\geq
v_0(k)$, with
\begin{equation}\label{vck}
v_0(k)=\frac{D_R+(D_p-D)k^2}{\sqrt{2}k}\;.
\end{equation}
The crossover from diffusive to propagating density fluctuations
is displayed qualitatively in Fig.~\ref{sound}. There is a lower
value $v_{c0}=\sqrt{2D_{R}\left( D_p-D\right)
}\sim\sqrt{D_{R}D_{0}}/2$ of $v_0$ below which the behavior is
always diffusive For a fixed value of $v_0>v_{c0}$ propagating
waves exist in a range of wavevector
 $k_{c1}\leq k\leq k_{c2}$, with
\begin{equation}
k_{c1,c2}=\frac{v_0}{D_p-D}\left(1\mp\sqrt{1-\frac{v_{c0}^2}{2v_0^2}}\right)\;.
\end{equation}
This range widens as $v_0$ increases, with
$\lim_{v_0\rightarrow\infty}k_{c1}=0$ and
$\lim_{v_0\rightarrow\infty}k_{c2}=\infty$.  This behavior
resembles closely  the appearance of propagating sound waves at
intermediate wavevectors in a compressible fluid that interacts
frictionally with a substrate \cite{Ramaswamy_Mazenko1982}. The
polarization here plays the same role as the flow velocity in the
Navier-Stokes equations of Re.~\cite{Ramaswamy_Mazenko1982} and
the self-propulsion effectively lowers the damping of polarization
fluctuations, yielding propagating waves.

Using $D_0\sim \ell^2 D_R$, we obtain $v_{c0}\sim D_0/\ell\sim
\ell D_R$. It is useful to introduce two characteristic length
scales as
\begin{eqnarray}
&&\ell _{diff}=\sqrt{\frac{2D}{D_{R}}}\;,\nonumber\\
&&\ell _{sp}=\frac{v_{0}}{D_{R}}\;, \end{eqnarray}
where $\ell_{diff}$ and $\ell_{sp}$ are the distances travelled by
the center of mass of a rod due to diffusion and self propulsion,
respectively, during the characteristic time scale for rotational
diffusion, $D_R^{-1}$.  The condition $v_0>>v_{c0}$ of large self
propulsion corresponds to $\ell_{sp}>>\ell_{diff}$. In this limit
\begin{eqnarray}
&&k_{c1}\simeq \frac{1}{\sqrt{2}}\frac{1}{\ell
_{sp}}+\frac{2\sqrt{2}}{\ell _{diff}}\left( \frac{\ell _{sp}}{\ell
_{diff} }\right)\;,\\
&&k_{c2}\simeq \frac{1}{\ell_{sp}}\;.
\end{eqnarray}
The modes are propagating for all length scales larger than
$\ell^2_{diff}/\ell_{sp}<<\ell_{diff}$ and smaller than
$\ell_{sp}$. For any fixed value of $v_0$, the modes always become
diffusive when $k\rightarrow 0$. Finally, at high density we
expect $D>D_p$. In this case propagating waves will exist in a
range of wavevector for all nonzero values of $v_0$. Furthermore
increasing the density will result to a strong suppression of
rotational diffusion from entanglement. In this case propagating
waves will exist for essentially all length scales.

\subsection{Fluctuations in the Nematic State}

Here we examine the linear stability of the nematic state,
characterized by a uniform density $\rho _{0}>\rho _{N}$,
alignment tensor $Q_{\alpha \beta }^{0}=S(\widehat{n}_{0\alpha
}\widehat{n}_{0\beta }-\frac{1}{2}\delta _{\alpha \beta })$, and
$\mathbf{P}=0$. The ordered state
 is symmetric for
$\widehat{\mathbf{n}}_{0}\rightarrow -\widehat{\mathbf{n}}_{0}$.
The hydrodynamic variables in this case are the density and the
director. As for the isotropic state, fluctuations in the
polarization, although overdamped, can qualitatively change the
system dynamics and ultimately render the uniform nematic state
unstable. They will therefore be incorporated in the analysis
below. We assume the system to be deep in the nematic state and
neglect fluctuations in the magnitude $S$ of the order parameter.
We consider small fluctuations about the ordered state by letting
\begin{eqnarray}
&&\rho \left( \mathbf{r},t\right) =\rho _{0}+\delta \rho \left( \mathbf{r}%
,t\right) \;,  \nonumber \\
&&\mathbf{P}\left( \mathbf{r},t\right) =\delta \mathbf{P}\left( \mathbf{r}%
,t\right) \;,  \nonumber \\
&&Q_{\alpha \beta }\left( \mathbf{r},t\right) =Q_{\alpha \beta }^{0}+S\big[%
\widehat{n}_{0\alpha }\delta n_{\bot \beta }\left(
\mathbf{r},t\right)
\nonumber \\
&&+\delta n_{\bot \alpha }\left( \mathbf{r},t\right) \widehat{n}_{0\beta }%
\big]\;,  \label{fluctuations}
\end{eqnarray}
where $\delta \mathbf{n}_{\bot }$ denotes fluctuations perpendicular to $%
\widehat{\mathbf{n}}_{0}$. To linear order, this is the only
fluctuation in the director field that preserves $|\widehat{\bf
n}|=1$. It is convenient to choose a coordinate system with the
$x$ axis along $\widehat{\mathbf{n}}_{0}$, so that $\delta
\mathbf{n}_{\bot }=\delta n_{y} \mathbf{\hat{y}}$.

\begin{figure}
\centerline{\epsfxsize=6cm \epsfbox{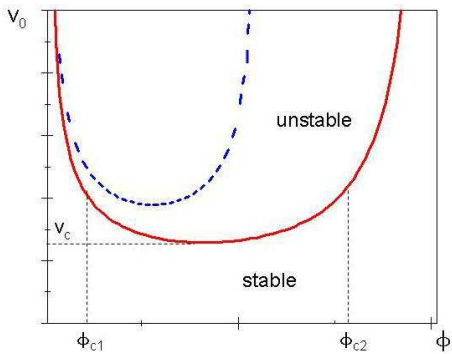}} \caption{(color
online) A schematic plot of the boundary $v_c(\phi,S)$ above which
the homogeneous nematic state is linearly unstable as a function
of $\phi$ for $S=1$ (solid red line) and $S=0.75$ (dashed blue
line). For a fixed value of $v_0>v_c(S)$ the nematic state is
unstable for $\phi_{c1}\leq\phi\leq\phi_{c2}$.} \label{alphac}
\end{figure}

The fluctuations $\delta \mathbf{P}$ in the polarization are
overdamped at the rate $D_{R}$. For $t>>D_{R}^{-1}$ we neglect
$\partial _{t}\delta \mathbf{P}$ and eliminate polarization
fluctuations in favor of the density and director fields. The
resulting hydrodynamic equations are given by
\begin{widetext}
\begin{eqnarray}
&&\partial _{t}\delta \rho =\Big(D_{x}\partial _{x}^{2}+D_{\perp
}\nabla _{\perp }^{2}\Big)\delta \rho +D_{0}S\rho _{0}\left(
1+2\alpha \right) \partial
_{x}\bm\nabla _{\perp }\cdot \delta \mathbf{n}_{\perp }\;, \\
&&\partial \delta \mathbf{n}_{\perp }=\Big(K_{3}\partial _{x}^{2}+K_{1}\bm%
\nabla _{\perp }\bm\nabla _{\perp }\cdot \Big)\delta \mathbf{n}_{\perp }+%
\frac{D_{0}}{8S}\left( 3+2\alpha \right) \partial _{x}\bm\nabla _{\perp }%
\frac{\delta \rho }{\rho _{0}}\;,
\end{eqnarray}
\end{widetext}
with
\begin{eqnarray}
&&D_{x}=\frac{D_{0}}{4}\left[3+S+2\alpha \left( 1+S\right)
\right] ,
\label{diff1} \\
&&D_{\perp }=\frac{D_{0}}{4}\left[3-S+2\alpha \left( 1-S\right)
\right] \;,
\label{diff2} \\
&&K_{1}=K_{3}=\frac{D_{0}}{4}(3+\alpha ),
\end{eqnarray}
and $\alpha =\frac{v_{0}^{2}}{D_{R}D}\sim
\frac{v_{0}^{2}}{4v_{c0}^{2}}$ a dimensionless parameter.
 As expected on the basis of
symmetry, only the square of the self propulsion velocity enters
the equations for the nematic. Self propulsion enhances diffusion
along the direction $x$ of alignment. It also stiffens both the
bend and splay elastic constants, $K_{3}$ and $K_{1}$. To analyze
the stability of the nematic state we expand the fluctuations in
Fourier modes at wavevector $\mathbf{k}$. Denoting by $\phi $ the
angle that $\mathbf{k}$ makes with $\widehat{\bf n}_0$, the
equations for the Fourier amplitudes of the fluctuations are
\begin{widetext}
\begin{eqnarray}
&&\partial _{t}\delta \widetilde{\rho }=-D_{\rho \rho }\left(
\alpha ,\phi \right) k^{2}\delta \widetilde{\rho }-S\rho
_{0}D_{\rho n}(\alpha ,\phi
)k^{2}\delta \widetilde{n}_{y}\;,  \label{2.3.4} \\
&&\partial _{t}\delta \widetilde{n}_{y}=-D_{n\rho }(\alpha ,\phi )k^{2}\frac{%
\delta \widetilde{\rho }}{S\rho _{0}}-D_{nn}(\alpha )k^{2}\delta
\widetilde{n}_{y}\;,  \label{2.3.5}
\end{eqnarray}
\end{widetext}
with
\begin{eqnarray}
D_{\rho \rho } &=&\left( D_{x}\sin ^{2}\phi +D_{\perp }\cos
^{2}\phi \right)
\;, \\
D_{nn} &=&\frac{D_{0}}{4}(3+\alpha )\;, \\
D_{\rho n} &=&D_{0}\left( 1+2\alpha \right) \cos \phi \sin \phi \;, \\
D_{n\rho } &=&\frac{D_{0}}{8}\left( 3+2\alpha \right) \cos \phi
\sin \phi
\end{eqnarray}
When $\phi =0$, i.e., for wavevectors $\mathbf{k}$  parallel to
$\widehat{\bf n}_0$, density and director fluctuations decouple
and the modes are always stable. To linear order, this geometry
corresponds to pure bend fluctuations of the director. The modes
also decouple for $\phi =\pi /2$, corresponding to $\mathbf{k}$
normal to the direction of broken symmetry. To analyze the linear
stability of the nematic state for arbitrary angles $\phi $ we
must examine the the dispersion relations of the hydrodynamic
modes associated with Eqs.~(\ref{2.3.4}) and (\ref{2.3.5}). These
are easily obtained as
\begin{equation}
\lambda _{\pm}=\frac{k^2}{2}\left[-(D_{\rho\rho}+D_{nn})\pm \sqrt{%
(D_{\rho\rho}-D_{nn})^2+4D_{\rho n}D_{n\rho}} \right] \;.
\label{2.3.6}
\end{equation}
The eigenvalues are always real, corresponding to diffusive modes. The mode $%
\lambda_-$ is always negative, indicating stable decay of
fluctuations. The mode $\lambda_+$ changes sign for
\begin{equation}
D_{\rho n}D_{n\rho }>D_{\rho \rho }D_{nn}\;.  \label{instab1}
\end{equation}
It is easy to verify that this condition can never be satisfied
when $v_{0} =0$, i.e., the nematic state is stable in the absence
of self propulsion. On the other hand,  the homogeneous nematic
state becomes unstable for all values of $k$ at large enough
$v_{0}$.  The condition for the onset of the instability obtained
from Eq.~(\ref{instab1}) yields a boundary $v_c(\phi,S)$. The
nematic state is unstable for $v_0>v_c(\phi,S)$, as shown in
Fig.~(\ref{alphac}). For a fixed value of $S$, there is a  minimum
threshold value $v_c(S)$ required to destabilize the nematic
state.  For a fixed value of $v_0>v_c(S)$ the nematic state is
unstable for all spatial gradients such that
$\phi_{c1}\leq\phi\leq\phi_{c2}$. For $S=1$, we find
$\lim_{v_0\rightarrow\infty}\phi _{c1}= \frac{\pi }{4}$ $\ $and
$\lim_{v_0\rightarrow\infty}\phi _{c2}=\frac{\pi }{2}$, and the
system becomes unstable to all spatial gradients such that the
splay wins over bend. Finally, note that the unstable region
shrinks in size as the magnitude of nematic order in the system
decreases.

The instability of the nematic state arises from a subtle
interplay of splay and bend deformations and diffusion
longitudinal and transverse to the direction of nematic order. It
only occurs for angles $\phi>\pi/4$, corresponding to situations
where splay deformations of the nematic director exceed bend
deformations. It does not, however, occur for pure splay
($\phi=\pi/2$), other than at $v_0\rightarrow\infty$. In  a
nematic, curvature inhomogeneities of the director  yield mass
currents, with splay deformations $\sim\partial_y\delta n_y$
yielding current along the direction $x$  of mean order and bend
deformations $\sim\partial_x\delta n_y$ yielding current along the
$y$ direction. Self propulsion enhances both contributions. In
addition, it enhances mass diffusion both longitudinal and
transverse to the direction of order, as indicated in
Eqs.~(\ref{diff1}) and (\ref{diff2}). However, while the
enhancement of $D_x$ from self propulsion  grows as the nematic
order increases and is maximum at $S=1$, the enhancement of
$D_\perp$ drops as one moves deeper into the nematic state and
vanishes when $S=1$. Thus, if a curvature in the director field
induces a mass flux that requires relaxation through appreciable
diffusion perpendicular to the mean director, then, beyond the
threshold value $v_{c}$, the system becomes unstable. A
fluctuation in the director field  corresponding to a large splay
and a small bend results in exactly such mass fluxes. Conversely,
when bend exceeds splay longitudinal diffusion is the main
mechanism that restores orientational order. This is enhanced by
self-propulsion and the system remains linearly stable.

The analysis described above only includes terms up to quadratic
order in the gradients and predicts that the nematic state is
unstable on all length scales. We expect that terms of higher
order in the gradients arising from excluded volume and diffusion
currents will stabilize the nematic state at large wavevectors
\cite{Marchetti06}, setting a length scale for the typical size of
the ordered regions, not unlike what has been observed in the
simulations by Peruani et al.\cite{Peruani06}

\section{Summary and Discussion}

In this paper we have discussed the collective dynamics of  self
propelled hard rods on a frictional substrate. The rods interact
through excluded volume interactions. The self propulsion was
implemented as a center of mass force acting on each rod that
propels it along  its long axis. Self-propulsion breaks the
nematic symmetry  at the level of the microscopic equations of
motion, but since it yields only a center of mass force on each
rod it is insufficient to generate a macroscopic  polarized state.
The existence of a polar state requires either microscopic torques
that turn the particle towards an externally determined direction
(for instance as seen in the case of chemotaxis in bacterial
motion) or an aligning interaction that can distinguish the two
ends of the extended object, which a physical excluded volume
interaction of rods does not. As a result, the only possible bulk
states of the self-propelled systems are those possible in
equilibrium : isotropic and nematic.

Self-propulsion, does, however have a profound effect on the
nature of the  fluctuations in each of this states. In the
isotropic state self-propulsion yields the appearance of
propagating waves in a range of wavevectors, as shown in
Fig.~\ref{sound}. This result is closely analogous to the
appearance of sound waves in a compressible fluid on a frictional
substrate, predicted many years ago by Ramaswamy and Mazenko
\cite{Ramaswamy_Mazenko1982}. This phenomenon has not been
observed in equilibrium systems where it requires a very small
value of the friction. In self-propelled systems, self-propulsion
itself effectively lowers the friction and should yield a wide range of
parameters where propagating density waves may be observable.
Propagating density waves may indeed have been seen in the
collective dynamics of epithelial cells on a substrate
\cite{Angelini}. In the nematic state fluctuations in the local
polarization yield mass fluxes and can destabilize the
homogeneous nematic state. This has been previously observed in
the numerical study of \cite{Peruani06}. Further, before the onset
of the instability, these same director fluctuations and the
resulting  anomalous mass flux associated with them account for
the large number fluctuations predicted in \cite{RamaswamyEPL}.

Our work has several limitations. First, we neglect all
correlations in deriving the Smoluchowski equation and in the
functional assumption made to obtain the hydrodynamic equations.
On the other hand, our hydrodynamic equations have precisely the
structure predicted on the basis of pure symmetry considerations.
This suggests that the behavior obtained here may be generic.
Therefore, further insight into the transport processes can be
obtained from numerical simulation of the model system and
comparing the results to the inherently low density theory
presented here. Secondly, the overdamped Langevin microdynamics
that describes the hard core interaction as an equilibrium
mean-field excluded volume effect may not be an adequate starting
point to incorporate self-propulsion. It can be shown that the
momentum transfer that occurs in the interactions between
self-propelled hard rods modifies the diffusion processes in the
system \cite{aparna2}. These effects can be captured by starting
with the true Langevin equations describing a system of hard rods
on a substrate and then obtaining the associated Fokker-Planck
equation, as will be described in a future
publication~\cite{aparna2}.

\section{Acknowledgement}

Research supported by the NSF through grants DMR-0305407 and
DMR-0705105. MCM also received support from the Institut Curie in
Paris via a  Rotschild-Yvette-Mayent sabbatical fellowship. She
thanks both the Institut Curie and ESPCI for their hospitality
during the completion of some of this work. Finally, we thank
Sriram Ramaswamy for useful discussions and for alerting us to
Ref.~\cite{Ramaswamy_Mazenko1982}.
\appendix
\section{Hydrodynamic Fluxes}
In this appendix, we give the expression for the excluded volume contributions to the
translational and rotational fluxes defined in Eqs.~(\ref{2.7}) and (\ref{2.8}). These have been obtained assuming a low moments
closure for the one-particle distribution function  of the form given in Eq.~(\ref{2.10}).
The excluded volume contributions to the translational fluxes are given by
\begin{widetext}
\begin{eqnarray}
J_{\alpha }^{ex} &=&D_{0}\eta _{0}\partial _{\beta
}[-\frac{1}{2}\rho ^{2}\left( Q_{\alpha \beta }+\frac{3}{4}\delta
_{\alpha \beta }\right) +
\frac{2}{9}\rho ^{2}Q_{\alpha \beta }^{2} +\frac{7}{18}\rho ^{2}\delta _{\alpha \beta
}TrQ^{2}]+\frac{2}{3}D_{0}\eta
_{0}\rho \partial _{\beta }\rho Q_{\alpha \beta } \;, \label{C.3}
\end{eqnarray}
\begin{eqnarray}
J_{\alpha \beta }^{ex} &=&-\frac{1}{8}D_{0}\eta
_{0}\left( \Delta _{\alpha \beta \gamma \sigma
}+4\delta _{\alpha \sigma }\delta _{\beta
\gamma }\right) \rho P_{\gamma }\partial _{\sigma }\rho  +\frac{1}{18}D_{0}\eta _{0}\left( \Delta _{\alpha \beta \omega
\gamma }+6\delta _{\alpha \omega }\delta _{\beta \gamma }\right)
\rho P_{\sigma
}\partial _{\omega }\rho Q_{\sigma \gamma }  \nonumber \\
&&+\frac{1}{18}D_{0}\eta _{0}[\left( \Delta _{\alpha \beta \omega
\gamma }-\delta _{\alpha \beta }\delta _{\omega \gamma }\right)
\rho P_{\omega
}\partial _{\sigma }\rho Q_{\sigma \gamma } +\rho P_{\omega }\partial _{\omega }\rho Q_{\alpha \beta }] \;, \label{C.7}
\end{eqnarray}
\begin{eqnarray}
J_{\alpha \beta \gamma }^{ex} &=&\frac{1}{12}D_{0}\eta _{0}\{-\frac{3}{8}%
\left( \delta _{\alpha \gamma }\delta _{\beta \sigma }+\delta
_{\alpha \sigma }\delta _{\beta \gamma }-\delta _{\alpha \beta
}\delta _{\gamma
\sigma }\right) \partial _{\sigma }\rho ^{2}  \nonumber \\
&&+\frac{1}{3}\rho \partial _{\sigma }\rho \left( 7\delta _{\gamma
\sigma }Q_{\alpha \beta }+\delta _{\alpha \gamma }Q_{\beta \sigma
}+\delta _{\beta \sigma }Q_{\alpha \gamma }+\delta _{\beta \gamma
}Q_{\alpha \sigma }+\delta _{\alpha \sigma }Q_{\beta \gamma
}-2\delta _{\alpha \beta }Q_{\gamma \sigma
}\right)   \nonumber \\
&&-\rho \left( 7\delta _{\gamma \sigma }Q_{\alpha \beta }+\delta
_{\alpha \gamma }Q_{\beta \sigma }+\delta _{\beta \sigma
}Q_{\alpha \gamma }+\delta _{\beta \gamma }Q_{\alpha \sigma
}+\delta _{\alpha \sigma }Q_{\beta \gamma }-2\delta _{\alpha \beta
}Q_{\gamma \sigma }\right) \partial _{\sigma }\rho
\nonumber \\
&&+\frac{1}{3}\partial _{\sigma }\rho ^{2}\left( Q_{\alpha \beta
}Q_{\gamma \sigma }+Q_{\alpha \gamma }Q_{\beta \sigma }+Q_{\alpha
\sigma }Q_{\gamma \beta }-\delta _{\alpha \beta }Q_{\gamma \omega
}Q_{\sigma \omega }\right)
\nonumber \\
&&+\rho ^{2}\left( \delta _{\alpha \sigma }Q_{\beta \omega
}+\delta _{\beta \sigma }Q_{\alpha \omega }-\delta _{\alpha \beta
}Q_{\sigma \omega }\right) Q_{\gamma \omega }+\rho ^{2}\left(
\delta _{\alpha \gamma }Q_{\beta \omega }+\delta _{\beta \gamma
}Q_{\alpha \omega }-\delta _{\alpha \beta }Q_{\gamma
\omega }\right) Q_{\sigma \omega }  \nonumber \\
&&+\frac{1}{4}\rho ^{2}\left( \delta _{\alpha \gamma }\delta
_{\beta \sigma }+\delta _{\alpha \sigma }\delta _{\beta \gamma
}-19\delta _{\alpha \beta }\delta _{\gamma \sigma }\right)
Tr\left( Q^{2}\right) +9\rho ^{2}\delta _{\gamma \sigma }Q_{\alpha
\beta }^{2} \}\;, \label{C.13}
\end{eqnarray}
\end{widetext}
where the notation
\begin{equation}
\Delta _{\alpha \beta \gamma \sigma }= \delta _{\alpha \beta
}\delta _{\gamma \sigma }+\delta _{\alpha \gamma }\delta _{\beta
\sigma }+\delta _{\alpha \sigma }\delta _{\beta \gamma }
\end{equation}
has been introduced for compactness.
The  excluded volume contributions to the rotational fluxes are
\begin{widetext}
\begin{eqnarray}
R_{\alpha }^{ex} &=&-\frac{1}{3}D_{R}\eta _{0}[4\rho ^{2}P_{\beta
}Q_{\alpha
\beta }+\frac{l^{2}}{6}\rho P_{\alpha }\nabla ^{2}\rho  -\frac{l^{2}}{3}\rho P_{\beta }\partial _{\alpha }\partial _{\beta }\rho +%
\frac{l^{2}}{18}\rho P_{\beta }\nabla ^{2}\left( \rho Q_{\alpha
\beta
}\right) ]  \nonumber \\
&&-D_{R}\eta _{0}\frac{l^{2}}{27}[\rho P_{\gamma }\partial
_{\gamma
}\partial _{\beta }\left( \rho Q_{\alpha \beta }\right)   +\left( \rho P_{\beta }\partial _{\alpha }\partial _{\gamma
}-\rho P_{\alpha }\partial _{\beta }\partial _{\gamma }\right)
\left( \rho Q_{\beta \gamma }\right) ]\;,
\end{eqnarray}
\begin{eqnarray}
R_{\alpha \beta }^{ex} &=&-\frac{4}{3}D_{R}\eta _{0}\rho
^{2}Q_{\alpha \beta }-\frac{1}{288}D_{R}l^{2}\eta _{0}\rho \left(
\partial _{\alpha }\partial _{\beta }\rho -\frac{1}{2}\delta
_{\alpha \beta }\nabla ^{2}\rho \right)
\nonumber \\
&&-\frac{1}{288}D_{R}l^{2}\eta _{0}\rho \left( \delta _{\alpha
\sigma }Q_{\beta \gamma }+\delta _{\beta \sigma }Q_{\alpha \gamma
}-\delta _{\alpha \beta }Q_{\gamma \sigma }\right) \partial
_{\gamma }\partial _{\sigma }\rho\;.
\label{C.14}
\end{eqnarray}
\end{widetext}


\begin{references}

\bibitem{cross_review} M. C. Cross and P. C. Hohenberg, Reviews of
Modern Physics, {\bf 65}, 851 (1993).

\bibitem{Vicsek} T. Vicsek, A. Czirok, E. Ben-Jacob, I. Cohen and
O. Shochet, Phys. Rev. Lett. {\bf 75}, 1226 (1995).

\bibitem{TonerRev} J. Toner, Y. Yu and S. Ramaswamy, Ann. Phys. {\bf
318}, 170 (2005).

\bibitem{ChaikinLubensky} P. M. Chaikin and T. C. Lubensky, {\em Principles of condensed matter
physics},Cambridge ; New York, NY (1995).

\bibitem{deGrootMazur}  S. R. de Groot and P. Mazur, {\em Non-equilibrium
thermodynamics},New York, Interscience Publishers (1962).

\bibitem{Lubensky_book}
 N.D. Mermin and H. Wagner, Phys. Rev. Lett. {\bf 17}, 1133 (1966)

\bibitem{Chate1} G. Gregoire and H. Chate, Phys. Rev. Lett., {\bf
92}, 025702 (2004).

\bibitem{Chate2} H. Chate, F. Ginelli and R. Montagne, Phys. Rev.
Lett., {\bf 96}, 180602 (2006).

\bibitem{RamaswamyPRL} R. A. Simha, S. Ramaswamy, Phys. Rev.
Lett., {\bf 89}, 058101 (2002).

\bibitem{RamaswamyEPL}S. Ramaswamy, R. A. Simha and J. Toner,Europhys. Lett. {\bf 62}
 196 (2003).

\bibitem{RamaswamyJSTAT} V. Narayanan, N. Menon and S. Ramaswamy,
J. Stat. Mech., P01005 (2006).

\bibitem{AdBh}C. R. Kube, H. Zhang Adaptive Behavior, {\bf 2}, 189
(1993).

\bibitem{BoltzPap} E. Bertin, M. Droz and G. Gregoire, Phys. Rev.
E, {\bf 74}, 022101 (2006).

\bibitem{Peruani06}  F. Peruani, A. Deutsch and M. Bar, Phys. Rev. E {\bf 74}, 030904 (R) (2006).

\bibitem{Aranson}
I. S. Aranson and L. S. Tsimring, Phys. Rev. E. {\bf 71},
050901(R) (2005)

\bibitem{Igoshin1} O. A. Igoshin, R. Welch, D. Kaiser, PNAS, {\bf 101}, 4256
(2004).

\bibitem{Reynolds} C. W. Reynolds, ACM SIGGRAPH Computer Grathetaphics, {\bf 21},25
(1987).

\bibitem{aparna2} A. Baskaran and M. C. Marchetti, unpublished.

\bibitem{McLBook} J.A. McLennan , {\em Introduction to Nonequilibrium
Statistical Mechanics}, (Prentice-Hall, New Jersey, 1989).

\bibitem{BDKS} J. J. Brey, J. W. Dufty, C. S. Kim, and A. Santos, Phys.
Rev. E {\bf 58}, 4638 (1998).

\bibitem{EdDoiBook} M Doi, S. F Edwards, {\it The Theory of Polymer Dynamics}, Oxford University Press (1986).

\bibitem{Marchetti06}  A. Ahmadi, M. C. Marchetti, T. B. Liverpool, Phys. Rev. E {\bf 74}, 061913
(2006).

\bibitem{MR06} S. Mishra, S. Ramaswamy, Phys. Rev. Lett. {\bf 97}, 090602
(2006).

\bibitem{Ramaswamy_Mazenko1982}
S. Ramaswamy and G. F. Mazenko, Phys. Rev. A {\bf 26}, 1735
(1982).

\bibitem{Angelini}
T. Angelini, M. Marquez, and D. Weitz, MAR07 Meeting of The
American Physical Society, Denver CO, 2007.

\end{references}
\end{document}